\documentclass{llncs}

\usepackage[margin=1in]{geometry}  
\usepackage{graphicx}              
\usepackage{listings}              
\usepackage{color}                 
\usepackage{amsmath}               
\usepackage{amsfonts}              

\usepackage{wrapfig}
\usepackage{multirow}

\definecolor{lightgray}{rgb}{.9,.9,.9}
\definecolor{darkgray}{rgb}{.4,.4,.4}
\definecolor{purple}{rgb}{0.65, 0.12, 0.82}
\definecolor{darkgreen}{rgb}{0.0, 0.3, 0.0}
\definecolor{darkblue}{rgb}{0.0, 0.0, 0.3}

\lstdefinelanguage{JavaScript}{
  keywords={typeof, new, true, false, catch, function, return, null, catch, switch, var, if, in, while, do, else, case, break},
  keywordstyle=\color{darkblue}\bfseries,
  ndkeywords={class, export, boolean, throw, implements, import, this},
  ndkeywordstyle=\color{darkgray}\bfseries,
  identifierstyle=\color{black},
  sensitive=false,
  comment=[l]{//},
  morecomment=[s]{/*}{*/},
  commentstyle=\color{darkgreen},
  stringstyle=\color{grey},
  morestring=[b]',
  morestring=[b]"
}

\newcommand{\RR}{\mathbb{R}}      
\newcommand{\ZZ}{\mathbb{Z}}      

\begin{document}

\nocite{*}

\title{Squiggle - A Glyph Recognizer for Gesture Input}

\author{Jeremy Lee \inst{1}}
\institute{The Unorthodox Engineers\\ \email{jeremy@unorthodox.com.au}}

\maketitle

\begin{abstract}
  Squiggle is a template-based glyph recognizer in the lineage of ``\$1 Recognizer''\cite{wob} and ``Protractor''\cite{yi}. It seeks a good fit linear affine mapping between the input and template glyphs which are represented as a list of milestone points along the glyph path. The algorithm can recognize input glyphs invariant of rotation, scaling, skew, and reflection symmetries. In practice the algorithm is fast and robust enough to recognize user-generated glyphs as they are being drawn in real time, and to project `shadows' of the matching templates as feedback.
\end{abstract}

\section{Introduction} \setlength{\columnsep}{0.5cm}
\begin{wrapfigure}{r}{0.45\textwidth}
  \vspace{-1.2cm}
  \begin{center}
    \includegraphics[scale=0.35,bb=0 0 579 419]{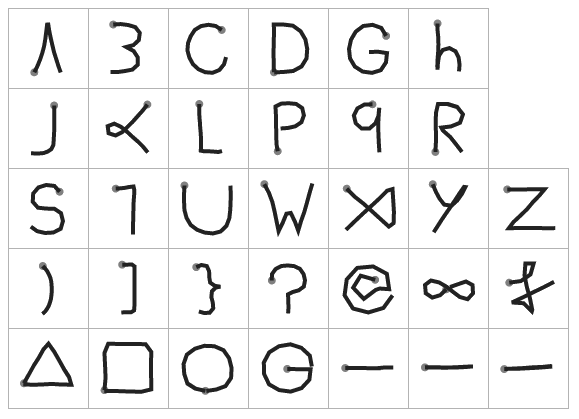}
  \end{center}
  \vspace{-0.5cm}
  \caption[]{The 33 templates of the Squiggle recognizer prototype, including glyphs that cause other template recognizers trouble such as the center-start circle and infinity symbols, and the single/double/triple-stroke lines.}
  \vspace{-0.5cm}
\end{wrapfigure}
Glyphs are the individual marks made during writing, from when the pen touches down to when it rises again. They can be as simple as the single dot above an ``i'', or as complex as an entire signature. Glyphs are the natural basic units involved in gestural computer input systems such as digitizer tablets, touch screens, and mouse input. The challenge for a glyph recognizer is to take the resulting digital path and guess what it \emph{means}. 

This is a deep problem: not only do we need to search the computational space using an appropriate numeric representation of the glyph, but our result has to match up with what a human being expects.

I present a novel algorithm called Squiggle which accepts an \emph{input glyph} and compares it against a library of \emph{template glyphs}. Glyphs are compared through an \emph{affine map} that projects the templates onto the input using simple linear matrix operations which can be performed efficiently on current hardware. The overlaid glyphs are geometrically compared, and the best match returned. The algorithm cleanly distinguishes between 0-D \emph{taps}, 1-D \emph{strokes} and full 2-D glyphs. Some orientations and reflections can optionally be excluded  per-glyph. 

\begin{figure}[b]
\vspace{-0.5cm}
\begin{center}
\begin{tabular}{cccccc}
\includegraphics[scale=0.38,bb=0 0 163 205]{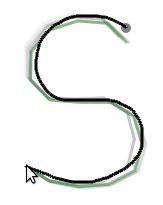} 
\includegraphics[scale=0.38,bb=0 0 181 228]{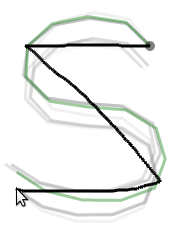} 
\includegraphics[scale=0.41,bb=0 0 194 202]{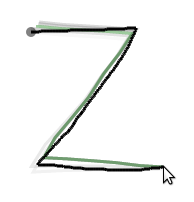} 
\includegraphics[scale=0.37,bb=0 0 206 237]{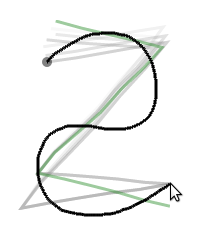} 
\includegraphics[scale=0.31,bb=0 0 215 275]{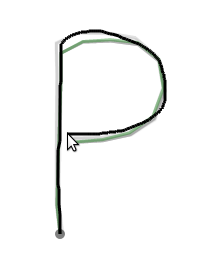} 
\includegraphics[scale=0.33,bb=0 0 178 258]{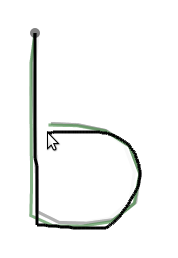}
\end{tabular}
\end{center}
\vspace{-0.5cm}
\caption{The prototype real-time glyph recognition software draws `shadows' of the best template alignment matches underneath the input pen path. Selective symmetry detection at early stages of the algorithm prevent `S' and `Z/N' domain overlap, but allows `P' and `b' to be recognized as one glyph despite rotation and reflection.}
\vspace{-0.5cm}
\end{figure}

The raw recognition rates of a template-based system are surprisingly irrelevant when its full capabilities are revealed to the user; to see and reprogram the templates. Success rates exceed 95\% on published data, but in practise can asymptote towards 100\% as the user adjusts templates that are consistently mis-recognized. Because the templates are projected onto the input glyph there is a natural ability to display the process to the user in their original geometry, which helps them learn the characteristics of the recognizer.

Out of the infinity of possible affine maps connecting the two spaces, we can use internal numeric features of the glyph geometry to quickly pare that down to just a handful of candidates. We depend on a couple of key insights:
\begin{itemize}
\item Affine maps between two co-ordinate systems can be defined by drawing a triangle (with labelled points) in each co-ordinate system, and then asserting they are the `same triangle'. 
\item The glyph path provides a series of points between which triangles can be constructed. No other points really matter, not even the origin.
\item We are not interested in comparing arbitrary triangles between glyphs, only the \emph{same logical triangle}. If we choose the first, eighth, and fourteenth points as indexes, that triangle exists in all glyphs with enough points.
\item If two glyphs are very similar, then choosing any logical triangle for the affine map will result in a good match.
\item The best triangles for the candidate affine map are the biggest, because vertex jitter will distort them the least. Conversely the smallest triangles are the worst, because they are often degenerate.
\end{itemize}

The combination of these insights is that we first analyze the input glyph to find its biggest internal triangles, then we pick out the matching logical triangles from the template glyphs for the other side of the affine map. These triangle pairs represent our best chance at a good alignment. If the glyphs are similar, this will project one straight onto the other, resulting in good scores. If they are not, this arbitrary choice will create a projection mess, and bad scores.
\vspace{-0.5cm}
\begin{center}
\begin{tabular}{p{5.5cm}p{1cm}p{8.5cm}}
\begin{center} \vspace{1cm}
\includegraphics[scale=0.4,bb=0 0 220 276]{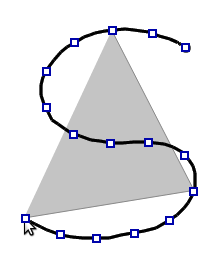} 
\end{center} 
& &
\begin{center}
\includegraphics[scale=0.20,bb=0 0 1132 812]{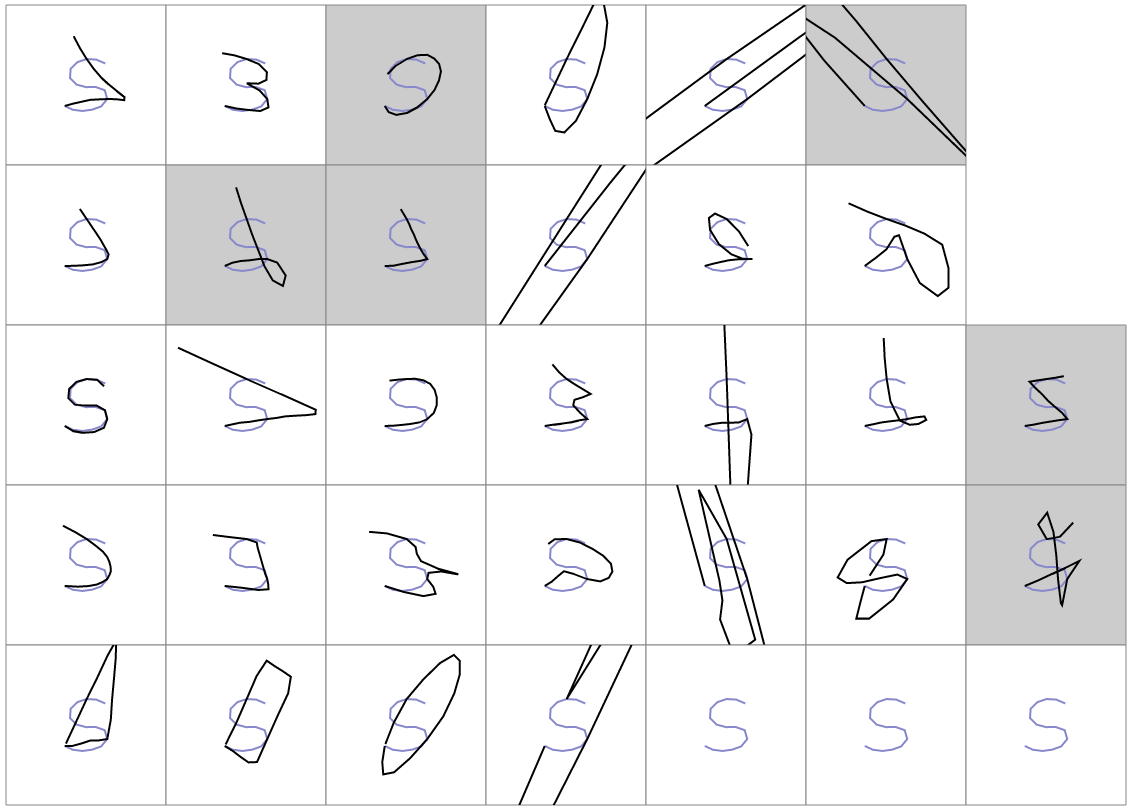}
\end{center} \vspace{-0.5cm} \\
\footnotesize{ \textbf{Fig.3.} The path is reduced to 16 milestone points, (equidistant along the original path) and it's largest internal triangles are found.} 
& &
\footnotesize{ \textbf{Fig.4.} Each template is projected onto the input glyph via affine transforms defined by the triangles. Some otherwise good matches (like 'Z') are excluded because they disallow reflections. The line glyphs are not even tested, because the input glyph is not degenerate.}
\end{tabular}
\end{center}
\setcounter{figure}{4}

We essentially \emph{assume} that the glyphs are identical and can be mapped directly onto each other through corresponding triangles. We perform the operation, and then see how badly it went. We don't attempt to correct bad assumptions, merely rank them. 

All the templates are therefore judged against the same assumption. Although it's possible for the wrong template glyph to win due to a poor choice of projection, it is surprisingly unlikely. (Just getting the 16 points in the right \emph{half} of a box by accident is a-priori $1:2^{16}$, as a drastic simplification)

If the glyphs are very similar, choosing a different triangle to perform the mapping will have very little effect: the affine map will be more or less the same. So if we have a good fit there is little point searching the triangles for a better one. If we have a bad fit, there is likewise little point broadening our criteria for one side of the map. Why bother? We're just going to blur the edges of the recognition domains. If finding a good fit is hard, it's obviously not the correct glyph.

The affine map intrinsically accounts for rotation, scaling, translation, skew, and reflection symmetries. (Although some symmetries such as rotation and reflection may be excluded by individual glyphs) It directly projects templates into `screen' co-ordinates, not an intermediate `normalized' co-ordinate system. This is important not just because the choice of co-ordinate system drastically affects the geometric difference metric; Points that are close together in one system may be stretched further apart by the affine map. Therefore it also makes sense to perform this comparison in the co-ordinate system that is `closest' to the user, to match their conception of distance and geometry.

\section{Algorithm Details}
The input glyph and template glyph paths are reduced to a fixed number of \emph{milestone} points along each path, typically 16, using the methods described in the previous papers. The two lists of points will be named $g$ for the input glyph and $h$ for the template glyph. Individual points from the path are nominated with a subscript index.
\begin{center}
\begin{tabular}{lrl}
$g = [ g_0 \dots g_{(n-1)} ]$ & $\qquad g_i$ & $\in \RR^2_g$ \\
$h = [ h_0 \dots h_{(n-1)} ]$ & $\qquad h_i$ & $\in \RR^2_h$ \\
& $n$ & $\approx 16$
\end{tabular}
\end{center}

Each of these sets of points is currently in its own co-ordinate system. Our task is to create an affine map $\widehat{A}$ which connects three points (equivalent to an origin point and two axis vectors, or the points of a triangle) across both co-ordinate systems.

It helps to know the total length $\lambda$ of the glyph path. If a direct record of the original path length is not available, you can just sum the segment lengths. This calculation is a lower bound, because these downsampled points are the milestones along a more detailed path that had curvature. Fortunately this number isn't critical, as long as it's a good approximation.
\begin{equation}
\lambda(p) \ge \sum_{i=0}^{n-2} \lvert p_{i+1} - p_i \rvert
\end{equation}

Next we construct our three-dimensional array of all possible triangles [a,b,c] (indexed by the choice of corner points) and fill each entry with a small 2x2 matrix made from the column vectors $(p_b-p_a)$ and $(p_c-p_a)$. We are interested only in the combinations (not permutations) of triangles made from the path points, so our 0-based indices are constrained to increase in order. 
\begin{equation}
M(p)_{abc} = [ p_b-p_a, p_c-p_a ] \qquad 0 \le a < b < c < n \in \ZZ
\end{equation}

We take the determinant of each 2x2 matrix, giving us twice the area of the triangle it represents. This is divided by the square of half the total glyph path length (the largest triangle that can be formed by a path is a 90-degree ``L'' shape with sides of half the total length) to give `normalized' determinant entries in a three dimensional matrix of scalars:
\begin{equation}
D(p)_{abc} = det(M(p)_{abc}) \div ( \frac{1}{2}\lambda(p) )^2
\end{equation}

This particular equation is so important it's worthwhile expanding it again and showing the full matrix form:
\begin{equation}
D(p)_{abc} = \frac{4}{\lambda(p)^2} \left\vert\left[\begin{matrix}
  p_b.x-p_a.x & \,\,\,\, & p_c.x-p_a.x \\[0.4em]
  p_b.y-p_a.y & \,\,\,\, & p_c.y-p_a.y
\end{matrix}\right]\right\vert
\end{equation}

Each triangle has therefore been given a score which expresses how large it is, relative to the scale of the entire glyph. `Small' and `large' triangles now have consistent absolute values, in the range 0...1. More usefully, we can choose a constant area $\epsilon$ below which triangles are considered `too degenerate' to be used as half of the affine map. This choice is not particularly critical, and experimentally was set to about $10^{-8}$. (Less than the size of a single pixel, if the glyph occupied the entire screen)

For the algorithm to proceed, we need to fully construct the \emph{normalized determinant matrix} $G_{abc}$ from the input glyph points $g_i$. If generated from 16 input points, this 3-D matrix will contain 560 entries. 
\begin{equation}
G_{abc} = D(g)_{abc}
\end{equation}

The matrix $H_{abc}$ is similarly constructed from the template glyph points $h_i$ and can either be stored with the template, or (since we only need the occasional sparse entry) we could just store the total path length and compute them on-the-fly. 
\begin{equation}
H_{abc} = D(h)_{abc}
\end{equation}

It's very useful to remember the maximum and minimum entry values of the input glyph matrix $G_{abc}$, as one of them corresponds to the top entry: the largest triangle. (Remember that negative values indicate mirror image triangles, so we need to pick the maximum absolute value in the matrix, not just the most positive.)
\begin{equation}
\top(G) = \underset{abc}{\text{MAX}}( abs( G ))
\end{equation}
\begin{equation}
[a,b,c] = triangle({ \top(G) })
\end{equation}

The above is a terrible abuse of notation, since of course the top entry and the triangle it corresponds to are found together during one iteration over the matrix, possibly during construction. Also, one should not believe a function that tries to turn a scalar value back into an index triplet. 

We use this `biggest triangle' result to make a critical decision: are we dealing with a line glyph, in which the entire glyph has essentially been drawn along a single line. (with allowances for wandering pen) Common examples are single \emph{strokes} or \emph{flicks} in various directions, or lines that double back. Any pointer input system needs to gracefully distinguish these 1-dimensional glyphs, just as they distinguish \emph{taps} or \emph{clicks} as a separate class of 0-dimensional glyph.

Line glyphs are problematic because almost any affine map built from one of its triangles will collapse the other glyph down to single line as well, resulting in false positive matches against glyph templates that were never designed to be used that way. Flatten a ``U'' enough, (or almost any character) and it looks just like a dash. The opposite problem happens when line glyphs are provided as templates: they can get artificially over-stretched so that a tiny kink in the middle might be exaggerated into the letter ``V''.

Note that there is no problem when comparing two glyphs which are \emph{both} line glyphs. in that case, degenerately collapsing one line onto another line is exactly what we expect, and the algorithm works correctly. It's only when you mix the classes that grief ensues.

Fortunately line glyphs are detected easily now we know the largest triangle that can be made from its constituent points. If it has very little area, then the glyph points are mostly in a single line. This is identical to the degeneracy test but with a much larger $\epsilon$ (about $0.002 \ldots 0.004$, compared with $10^{-8}$) which was settled on through experimentation and testing to gauge what humans consider `straight-ish'. 

Importantly, once the algorithm detects that it is in line glyph mode it stops caring about keeping degenerate triangles out of the affine map. We now \emph{expect} that both glyphs will contain mostly low-area degenerate triangles. This is the only difference between the modes; degenerate triangles on the template side of the map were a criteria for immediate rejection when matching against 2-D glyphs.

Note that only by treating the 1-D glyph as a 2-D glyph \emph{first} (trying to compute triangle areas) can we tell that it's not. This is in contrast with 0-D taps and clicks which are trivially self-evident.

In order to give us more than one chance to properly match up the glyph geometries, we select the biggest $m$ triangles from the matrix for testing. Various $m$ have been experimentally tried, and values from 8-16 seem to be perfectly adequate. The algorithm continues to function with only a partial loss of discrimination even if you reduce this to 2 or even 1! (Which scarcely seems possible, but if the glyphs are very similar then \emph{any} triangle will do)

Computationally, constructing the triangle determinant matrix takes about 1,680 MACs (Multiply-Accumulate operations) to turn the 16 point path into a normalized determinant matrix. Each geometric projection takes about 64 MACs, and the geometric distance test another 32 MACs, so testing with the ten biggest triangles ($m=10$) over a library of 32 template glyphs would consume 30,720 MACs. This is trivial on modern processors.

This neglects the \emph{sorting time} needed to find the biggest $m$ triangles in the input glyph matrix. Beware using a naive sorting algorithm (like shell sort) at this step, which could easily consume $\mathcal{O}(n^2)$ CAS-equivalent operations (313,600 for our 560 entry matrix) taking /emph{ten times longer than everything else combined}, just to return the top few triangles.

Since we don't need the result set in order, we can do a \emph{pivot select} where we pick a value, count how many matrix entries it would include and then adjust the pivot value until it selects just the $m$ entries we're after, which we collect on a final sweep. It takes about $\mathcal{O}(log \, n) = 9$ passes on average for our matrix, with 560 Compares per pass, so about the same number of read/compares as the merge sort: 5,020 on average) but \emph{no memory writes} until the final m are collected to the result list. Also each step compares the constant pivot value against each entry, rather than comparing two entries, so the \emph{read bandwidth} is halved. 

The pivot test selection can fail (or infinite loop) if there isn't a valid pivot value between the $m$th and $(m+1)$th ranked entries, such as when they're tied. An epsilon value stops the recursion when the range becomes too narrow. 

Now that we have our set of candidate triangles on the input glyph (selected because they give us the greatest \emph{chance} of matching against any template) we go through our library of template glyphs and perform the affine mapping of the template into the co-ordinate system of the input glyph so that the \emph{corresponding triangle} of the template is forced to fit exactly over the input triangle. 

This is the point at which we can do selective symmetry and orientation detection, to exclude undesirable matches. If the determinants of the input and template triangles are opposite in sign, then we are about to test glyphs invariant of mirror symmetry. 

Time to abuse some more tensor notation: $\widehat{p}_{abc}$ means the affine transformation matrix generated from the triangle with its points at indices $[a,b,c]$ on the path $p$. Each transform is a 3x2 matrix made from three column vectors built almost directly from the points of the triangle. 
\begin{equation}
\widehat{p}_{abc} = [p_b-p_a, \, p_c-p_a, \, p_a] \qquad 0 \le a < b < c < n \in \ZZ  
\end{equation}

Or if you prefer to see it expressed as the full matrix:
\begin{equation}
\widehat{p}_{abc} = \left[\begin{matrix}
  p_b.x-p_a.x & \,\,\,\, & p_c.x-p_a.x & \,\,\,\, & p_a.x  \\[0.4em]
  p_b.y-p_a.y & \,\,\,\, & p_c.y-p_a.y & \,\,\,\, & p_a.y
\end{matrix}\right]
\end{equation}

2-dimensional points to be run through the transform are put into a 3-vector row with 1 in the last entry and then multiplied by the transformation matrix to produce a column vector [x`, y`]. This is also equivalent to a standard vector / matrix multiplication if we pretend the 3x2 matrix is actually 3x3, with [0,0,1] in the bottom row, and we assume every vector has `1' in its third element.

The slightly older way of defining an affine transform was to use a 2x2 matrix plus a vector addition so that only 2-vectors had to be involved. I use the 3x2 construction because it is the standard form used by modern computer graphics software and hardware.

We will continue to use lower case letters with a single index to denote an individual point $p_i$ that exists in a path. We should also be distinguishing row and column vectors using something like bra-ket notation:
\begin{equation}
\left\langle p_i.x , p_i.y , 1 \right\vert \times \widehat{t}_{abc}  = \left\vert q_i.x , q_i.y \right\rangle
\end{equation}

Or we can just abuse the notation a little more for convenience and clarity and say, to indicate the transformation of a point:
\begin{equation}
p_i \times \widehat{t}_{abc} = q_i \qquad \qquad p_i \in \RR^2_p \qquad q_i \in \RR^2_q
\end{equation}

Since we generally transform entire paths at once, we could also say:
\begin{equation}
p \times \widehat{t}_{abc} = q
\end{equation}

Each transform has an inverse:
\begin{equation}
q \times \widehat{t}_{abc}^{-1} = p
\end{equation}

In this case we're not dealing with a complete 3x3 matrix, so the inverse is quite a bit easier to construct using Laplace's formula:
\begin{equation}
\left[\begin{matrix}
  a & c & e  \\[0.4em]
  b & d & f  \\[0.4em]
  0 & 0 & 1
\end{matrix}\right]^{-1} = \left[\begin{matrix}
  \frac{d}{ad-cb} & -\frac{c}{ad-cb} & \frac{cf-de}{ad-cb}  \\[0.4em]
  -\frac{b}{ad-cb} & \frac{a}{ad-cb} & -\frac{af-be}{ad-cb}  \\[0.4em]
  0 & -0 & 1
\end{matrix}\right]
\end{equation}

Now that we can turn triangles in each co-ordinate system into affine transforms, we can finally declare the \emph{affine map} $\widehat{A}$ for our choice of triangles by linking the two affine transformations together.
\begin{equation}
\widehat{A} : \, \widehat{h}_{abc} \, \mapsto \, \widehat{g}_{abc} \qquad \qquad g_i \in \RR^2_g \qquad h_i \in \RR^2_h
\end{equation}

We use this to project all the points $h_i$ into a new set of points $r_i$ in the co-ordinate system of $g$. These are the points we are going to `recognize' against the input glyph. Numerically we do this by multiplying $h_i$ by the matrix inverse of $\widehat{h}_{abc}$ to `leave' that co-ordinate system, and then multiplying by $\widehat{g}_{abc}$ to `enter' the new co-ordinate system to create a new set of points $r_i$.
\begin{equation}
h_i \times \widehat{h}_{abc}^{-1} \times \widehat{g}_{abc} = r_i \qquad \qquad h_i \in \RR^2_h \qquad r_i \in \RR^2_g
\end{equation}

If you are careful, you can pre-multiply the two matrices together and thereby perform the affine mapping in a single computational step.
\begin{equation}
\widehat{t}_{abc} = \widehat{h}_{abc}^{-1} \times \widehat{g}_{abc}
\end{equation}
\begin{equation}
h_i \times \widehat{t}_{abc} = r_i 
\end{equation}

We can drop the indexes and apply the transform to entire paths, in this case to turn the template into a \emph{recognizer path} to be compared against the input glyph:
\begin{equation}
h \times \widehat{t}_{abc} = r 
\end{equation}

The \emph{geometric distance} calculation we perform between the input and recognizer glyphs is a sum of the errors between the paired-off points.
\begin{equation}
metric(g,r) = \sum_{i=0}^{n-1}(g_i.x-r_i.x)^2 + (g_i.y-r_i.y)^2
\end{equation}
\vspace{-0.2cm}
\begin{figure}[h!]
\vspace{-0.5cm}
\begin{center}
\includegraphics[scale=0.25,bb=0 0 1748 1356]{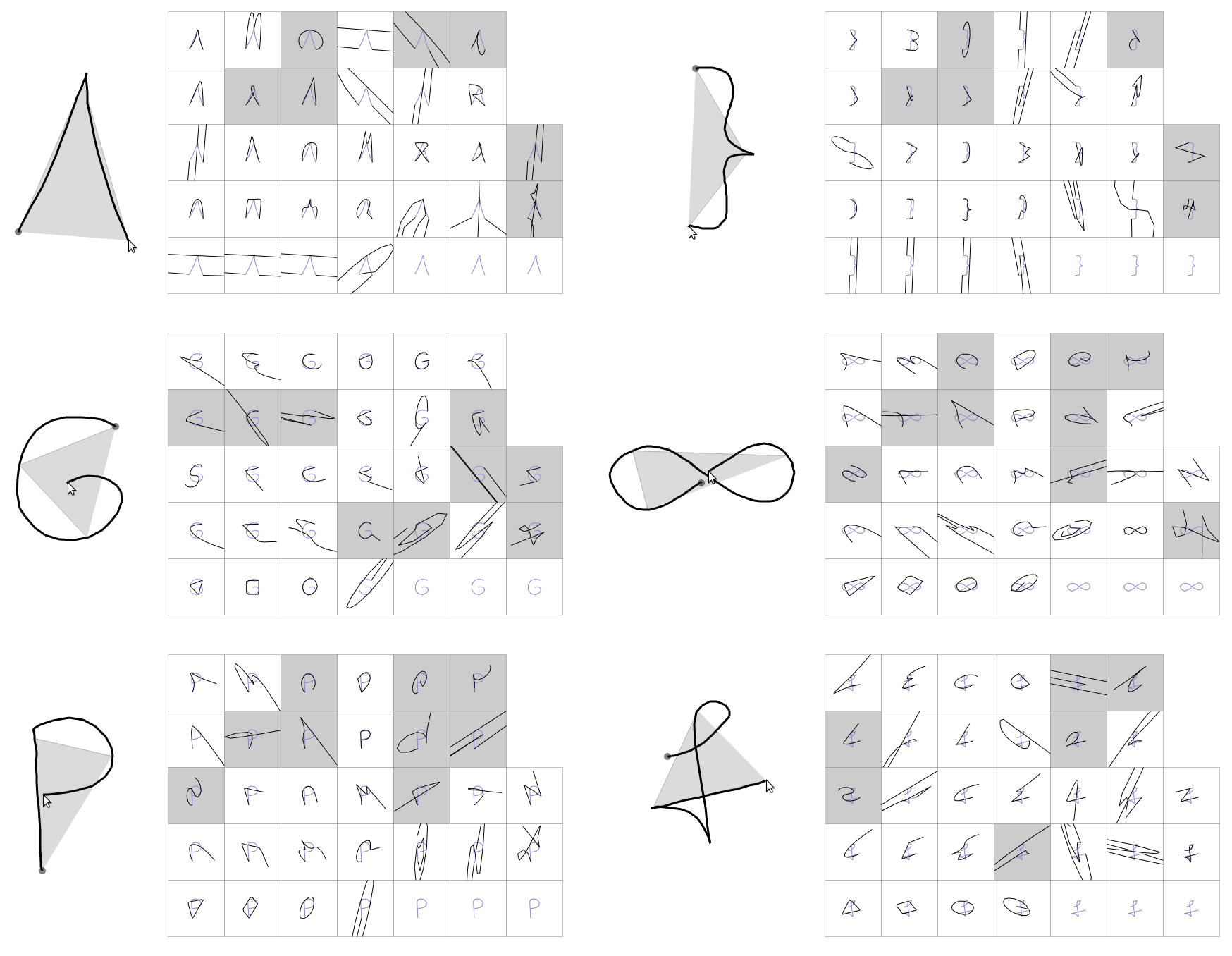}
\end{center}
\vspace{-0.5cm}
\caption{Input glyphs, their largest triangles, and the projections of template glyphs onto them.
Some individual templates (in grey) specifically disallow reflection symmetries, detected by triangle pairs which have determinants of opposite sign.}
\vspace{-0.5cm}
\end{figure}

Note that this is not quite the same as 2-D Euclidean distance, as we have omitted taking the square root. We are free to meddle with the metric, because there is no purely mathematical `best' function: we are trying to choose a behaviour that lines up with human expectations of path closeness, and this is more subtle than pure geometry.

\section{Orientation Detection}

In order to make some meaningful comparisons with the \$1 recognizer and Protractor algorithms, an `orientation detector' was added with the intention of limiting glyphs matches to within $45^{\circ}$ of their templates. 

Because of the affine mapping, Squiggle does not have a simple conception of `glyph angle', and there are no special orientation vectors that would remain undistorted for 2-D glyphs. Line glyphs do have a natural axis, but such glyphs are not handled by the other two algorithms.

To address this, a \emph{triangle similarity test} was invented which sums the cosine angles between the corresponding triangle edges of $g_{abc}$ and $h_{abc}$. 
\begin{equation}
\text{similarity}_{abc} = cos(\theta_{ab}) + cos(\theta_{bc})  + cos(\theta_{ca}) 
\end{equation}
where
\begin{equation}
cos(\theta_{mn}) = \frac{ (g_m - g_n) \cdot (h_m - h_n) }{ \lVert g_m - g_n \rVert \lVert h_m - h_n \rVert }
\end{equation}

This score ranges from ${-3}\ldots{3}$. A right isocoles triangle (with equal sides surrounding the right angle) rotated by $45^{\circ}$ produces a similarity score of 2.14. This rapidly drops to 0 at $90^{\circ}$, and then goes negative. This score is invariant to scaling, but is affected by skew.

\section{Experimental comparisons with \$1 and Protractor}
\begin{wrapfigure}{r}{0.45\textwidth}
  \vspace{-2.5cm}
  \begin{center}
    \includegraphics[scale=0.25,bb=0 0 810 777]{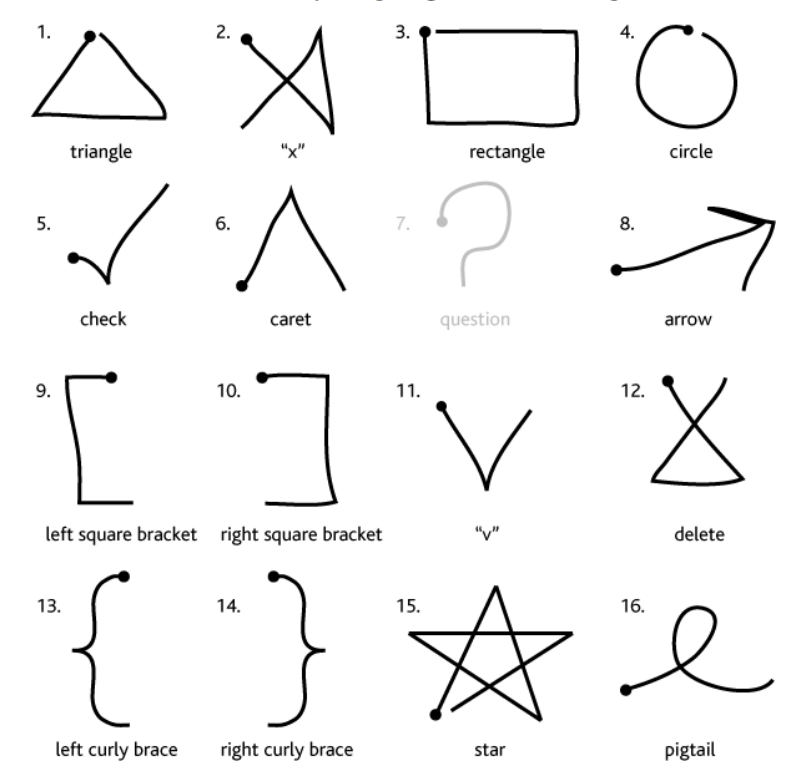}
  \end{center}
  \vspace{-0.5cm}
  \caption{The 16 example glyphs represented by the `\$1 Recognizer' data set. The question mark template data was not available, so it's gestures were not tested.}
  \vspace{-0.5cm}
\end{wrapfigure}
Wobbrock, Wilson and Li's excellent paper ``Gestures without Libraries, Toolkits or Training:  
A \$1 Recognizer for User Interface Prototypes'' provided much of the inspiration and framework for the development of Squiggle. Their published set of templates and input gesture data was used for comparative testing.

This data set comprises of 5280 input gestures representing 16 glyphs collected from 11 users at 3 different drawing speeds. The method of collection is described in detail in their paper, but can be summed up as: Eleven different subjects were presented with an image of the template and asked to redraw the glyph on a pen-based computer. Each user drew the glyphs ten times at three different speeds; fast, medium, and slow. Some of the gestures are quite `messy' and would be difficult even for a human being to correctly classify.

Since the original template data was not available, template glyphs were extracted from the code for their example web application. This created a small problem as one glyph (the question mark) had been replaced for the on-line version by a new `zig-zag' template. Therefore, both the zig-zag template and question mark gestures were omitted from the testing, giving 4950 input gestures for 15 template glyphs.

This data set is in many ways a worst case scenario for Squiggle in such a comparison. The low number of templates means Squiggle's high input glyph analysis costs are not amortized over a large set. The template set is quite `square', and does not contain any line glyphs which the other two algorithms are unable to handle. All glyphs begin and end far from the center, avoiding the singularity that would confuse their `indicative angle' calculation but to which Squiggle is immune. No use is made of mirror symmetries, and some glyphs are essentially affine transforms of others. Despite these disadvantages Squiggle performs admirably, producing an accuracy rate within 0.5\% of the \$1 Recognizer on this data.

The three recognizers were compared against the input data, and also against each other's conclusions. All three algorithms were run within the same execution environment from the same in-memory data set. Run times include only the core recognition processing time, and omit template setup and other start-up costs. Times were measured over five trials, and the three most consistent times were averaged.

The recognition accuracy rates, runtimes, and correlation between the outputs of these three algorithms are summarized in the tables below. The most important result is the near-parity in accuracy rates between \$1 and Squiggle.

\setlength{\columnsep}{1cm}
\begin{multicols}{2}

\begin{center}
\begin{tabular}{ l p{0.5cm} r p{0.5cm} r }
\multicolumn{5}{ c } { \footnotesize \textbf{Table 1.} Recognition accuracy and runtime}  \\
\multicolumn{5}{ c } { \footnotesize on the `\$1 Recognizer' gesture set.}  \\[6pt]
\hline
\footnotesize\textbf{Algorithm}  && \footnotesize\textbf{Accuracy} && \footnotesize\textbf{Runtime} \\
\hline
\$1 Recognizer && 95.56\% && 2.038s  \\
Squiggle   && 95.09\% && 3.292s  \\ 
Protractor && 92.87\% && 0.254s  \\ 
\hline
\end{tabular}
\end{center}

\begin{center}
\begin{tabular}{ l p{0.5cm} r  }
\multicolumn{3}{ c } { \footnotesize \textbf{Table 2.} Correlation of recognized glyphs }  \\
\multicolumn{3}{ c } { \footnotesize between pairs of algorithms.}  \\[6pt]
\hline
\footnotesize\textbf{Algorithm Pair} && \footnotesize\textbf{Correlation} \\ 
\hline
Protractor vs. \$1 Recognizer && 96.28\%  \\ 
Squiggle vs. \$1 Recognizer && 94.85\%  \\
Squiggle vs. Protractor && 92.77\%  \\ 
\hline
\end{tabular}
\end{center}

\end{multicols}
\setcounter{table}{2}

It is interesting to break down the recognition match counts for these two algorithms, to see where the mistakes are being made. The following table summarizes how each algorithm has classified the input gestures. There is a visible difference in the pattern of errors.

\begin{table}
\caption{Input gesture to glyph template recognition counts, for both the Squiggle and \$1 Recognizers.}
\resizebox{\textwidth}{!} {
\begin{tabular}{ l r p{0.1cm} r r r r r r r r r r r r r r r p{0.4cm} r r r r r r r r r r r r r r r l }
&&& \multicolumn{15}{c} { \footnotesize \textbf{Squiggle} } && \multicolumn{15}{c} { \footnotesize \textbf{\$1 Recognizer} } \\\footnotesize\textbf{Input Gesture} &&& \includegraphics[height=12pt,bb=0 0 170 170]{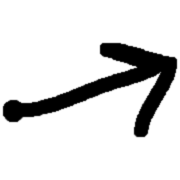} & \includegraphics[height=12pt,bb=0 0 170 170]{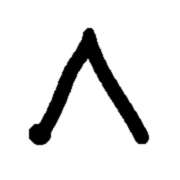} & \includegraphics[height=12pt,bb=0 0 170 170]{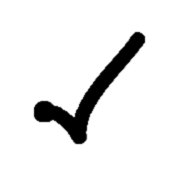} & \includegraphics[height=12pt,bb=0 0 170 170]{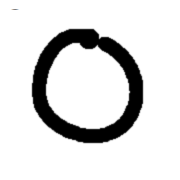} & \includegraphics[height=12pt,bb=0 0 170 170]{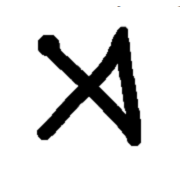} & \includegraphics[height=12pt,bb=0 0 170 170]{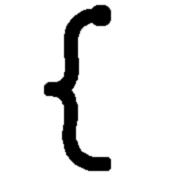} & \includegraphics[height=12pt,bb=0 0 170 170]{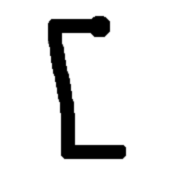} & \includegraphics[height=12pt,bb=0 0 170 170]{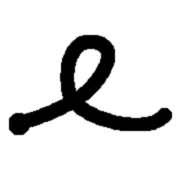} & \includegraphics[height=12pt,bb=0 0 170 170]{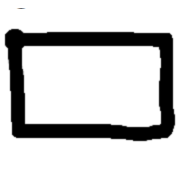} & \includegraphics[height=12pt,bb=0 0 170 170]{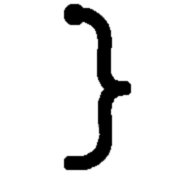} & \includegraphics[height=12pt,bb=0 0 170 170]{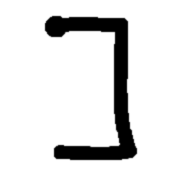} & \includegraphics[height=12pt,bb=0 0 170 170]{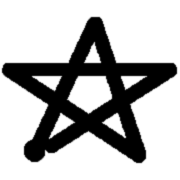} & \includegraphics[height=12pt,bb=0 0 170 170]{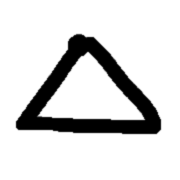} & \includegraphics[height=12pt,bb=0 0 170 170]{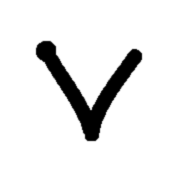} & \includegraphics[height=12pt,bb=0 0 170 170]{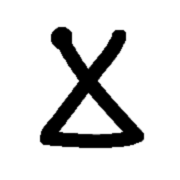} && \includegraphics[height=12pt,bb=0 0 170 170]{bold-arrow.png} & \includegraphics[height=12pt,bb=0 0 170 170]{bold-caret.png} & \includegraphics[height=12pt,bb=0 0 170 170]{bold-check.png} & \includegraphics[height=12pt,bb=0 0 170 170]{bold-circle.png} & \includegraphics[height=12pt,bb=0 0 170 170]{bold-delete_mark.png} & \includegraphics[height=12pt,bb=0 0 170 170]{bold-left_curly_brace.png} & \includegraphics[height=12pt,bb=0 0 170 170]{bold-left_sq_bracket.png} & \includegraphics[height=12pt,bb=0 0 170 170]{bold-pigtail.png} & \includegraphics[height=12pt,bb=0 0 170 170]{bold-rectangle.png} & \includegraphics[height=12pt,bb=0 0 170 170]{bold-right_curly_brace.png} & \includegraphics[height=12pt,bb=0 0 170 170]{bold-right_sq_bracket.png} & \includegraphics[height=12pt,bb=0 0 170 170]{bold-star.png} & \includegraphics[height=12pt,bb=0 0 170 170]{bold-triangle.png} & \includegraphics[height=12pt,bb=0 0 170 170]{bold-v.png} & \includegraphics[height=12pt,bb=0 0 170 170]{bold-x.png} &\\ \hline
arrow & \includegraphics[height=12pt,bb=0 0 170 170]{bold-arrow.png} && 285 & 44 & - & - & - & - & - & - & - & - & - & 1 & - & - & - && 328 & - & - & - & - & - & 1 & - & - & - & - & - & - & - & 1 &\\
caret & \includegraphics[height=12pt,bb=0 0 170 170]{bold-caret.png} && - & 330 & - & - & - & - & - & - & - & - & - & - & - & - & - && - & 330 & - & - & - & - & - & - & - & - & - & - & - & - & - &\\
check & \includegraphics[height=12pt,bb=0 0 170 170]{bold-check.png} && - & - & 304 & - & - & - & - & - & - & - & - & - & - & 26 & - && - & - & 326 & - & - & - & - & - & - & - & - & - & - & 4 & - &\\
circle & \includegraphics[height=12pt,bb=0 0 170 170]{bold-circle.png} && - & - & - & 305 & - & - & 4 & - & 5 & - & - & - & 15 & - & 1 && - & - & - & 315 & - & - & 1 & - & 14 & - & - & - & - & - & - &\\
delete mark & \includegraphics[height=12pt,bb=0 0 170 170]{bold-delete_mark.png} && - & - & - & - & 328 & - & - & - & - & - & - & - & - & 2 & - && - & - & - & - & 330 & - & - & - & - & - & - & - & - & - & - &\\
left curly & \includegraphics[height=12pt,bb=0 0 170 170]{bold-left_curly_brace.png} && - & - & - & - & - & 299 & 31 & - & - & - & - & - & - & - & - && - & - & 4 & - & - & 291 & 35 & - & - & - & - & - & - & - & - &\\
left bracket & \includegraphics[height=12pt,bb=0 0 170 170]{bold-left_sq_bracket.png} && - & - & - & - & - & 2 & 328 & - & - & - & - & - & - & - & - && - & - & - & - & - & - & 326 & - & 1 & - & - & - & - & 3 & - &\\
pigtail & \includegraphics[height=12pt,bb=0 0 170 170]{bold-pigtail.png} && 1 & 8 & - & - & - & - & - & 321 & - & - & - & - & - & - & - && 4 & 1 & - & - & - & - & - & 321 & - & - & - & - & - & - & 4 &\\
rectangle & \includegraphics[height=12pt,bb=0 0 170 170]{bold-rectangle.png} && - & - & - & 28 & - & - & - & - & 295 & - & - & - & 7 & - & - && - & - & - & 8 & - & - & - & - & 318 & - & - & - & 4 & - & - &\\
right curly & \includegraphics[height=12pt,bb=0 0 170 170]{bold-right_curly_brace.png} && - & - & - & - & - & - & - & - & - & 274 & 56 & - & - & - & - && - & - & - & - & - & - & - & - & - & 252 & 78 & - & - & - & - &\\
right bracket & \includegraphics[height=12pt,bb=0 0 170 170]{bold-right_sq_bracket.png} && - & - & - & - & - & - & - & - & - & - & 330 & - & - & - & - && - & 21 & - & - & - & - & - & - & - & - & 309 & - & - & - & - &\\
star & \includegraphics[height=12pt,bb=0 0 170 170]{bold-star.png} && - & - & - & - & - & - & 1 & - & - & - & - & 329 & - & - & - && - & - & - & - & - & - & - & - & - & - & - & 330 & - & - & - &\\
triangle & \includegraphics[height=12pt,bb=0 0 170 170]{bold-triangle.png} && - & - & - & 4 & - & - & 1 & - & - & - & - & - & 325 & - & - && - & - & - & - & - & - & - & - & 15 & - & - & - & 315 & - & - &\\
v & \includegraphics[height=12pt,bb=0 0 170 170]{bold-v.png} && - & - & 5 & - & - & - & - & - & - & - & - & - & - & 325 & - && - & - & 11 & - & - & - & 7 & - & - & - & - & - & - & 312 & - &\\
x & \includegraphics[height=12pt,bb=0 0 170 170]{bold-x.png} && - & - & - & - & - & - & 1 & - & - & - & - & - & - & - & 329 && 1 & - & 1 & - & - & - & - & 1 & - & - & - & - & - & - & 327 &\\
\hline
\end{tabular}
}
\end{table}

Squiggle consistently mis-recognizes arrows as carets, (but oddly never carets as arrows) checks as `v's, and rectangles as circles, whereas \$1 does not. This is explained by the affine mapping that Squiggle applies. These glyphs in particular can be transformed almost exactly onto each other through an appropriate affine map with significant skew, which is the symmetry that \$1 is \emph{not} invariant to.

The other interesting result is that Squiggle consistently mis-recognizes rectangles as circles but much less often circles as rectangles, perhaps indicating that subjects are `rounding their rectangles' more than vice-versa. (A brief examination of the data confirmed that users are indeed often drawing rectangles more like squares.)

Both algorithms have difficulty distinguishing the square and curly brackets as one might expect, but Squiggle performs this relatively hard task much better than \$1. In fact anywhere that a triangular gesture is compared against a square template, Squiggle almost never makes a mistake; in contrast to \$1 often mistaking carets for right brackets, and triangles for rectangles.

\pagebreak
\section{Code}
The following is not pseudocode, but the working JavaScript/ECMAScript used to implement the prototype system and perform tests. It is optimized for readability, and the semantics of the language are well defined. This approach should avoid transcription errors and be useful to implementers. The entire algorithm and all supporting functions is included, providing a complete example that hopefully avoids any ambiguity. 
\setlength{\columnsep}{0.1cm}
\begin{multicols}{2}

\begin{lstlisting}
// add a template to the recognizer
function add_template(name,path,mirror) {
	var g = { name: name, mirror: mirror };
	// regularize the path into segments 3 pixels long
	var rpath = path_regularize(path, 3);
	// interpolate the path down to 16 points
	g.path = path_interpolate(rpath.points, 16);
	// construct a normalized triangle determinant matrix
	g.ntm = path_ntm(g.path);
	// add template to the global array
	template_glyphs.push(g);
}

// recognize an input glyph
function recognize(path) {
	// regularize the path into segments 3 pixels long
	var rpath = path_regularize(path, 3);
	// is it long enough to process as a glyph?
	if(rpath.points!==undefined) if(rpath.points.length>4) {
		// interpolate the path down to 16 points
		var ipath = path_interpolate(rpath.points, 16);
		// construct a normalized triangle determinant matrix
		var ntm = path_ntm(ipath);
		// find the biggest 6 triangles +-2
		var pivot = ntm_pivot_for(ntm, 8, 2);
		var triangles = ntm_pivot_set(ntm, pivot);
		// var triangles = ntm_largest(ntm, 8);
		// find the best template match
		var g = { path: ipath, ntm: ntm };
		return match_glyph(g, triangles);
	}
}

// test a glyph against templates using alignment triangles
function match_glyph(g, align) {
	var best_glyph, best_index, best_score;
	for(var i in template_glyphs) {
		var h = template_glyphs[i];
		// are the glyphs the same line type?
		if(g.ntm.line===h.ntm.line) {
			// go through the alignment triangles
			for(var j in align) {
				var tri = align[j];	var test = true;
				// do we care about the triangle areas?
				if(!g.ntm.line) {
					// get the normalized determinants of the triangles
					var nd1 = ntm_entry(g.ntm,tri[0],tri[1],tri[2]);
					var nd2 = ntm_entry(h.ntm,tri[0],tri[1],tri[2]);
					// is either essentially zero?
					if( (Math.abs(nd1)<0.00000001) || 
						(Math.abs(nd2)<0.00000001) ) test = false;
					// exclude mirror symmetry?
					if(((nd1*nd2)<0) && (!h.mirror)) test = false;
					// optional orientation test
					test = test && tri_similarity(g.path,h.path,tri) > 2.12;
				}
				if(test) {
					// project the glyph into alignment
					var r = path_project(g.path, h.path, tri);
					// compute an error metric between the paths
					var metric = 0;
					for(var k in r) {
						var p1 = g.path[k]; var p2 = r[k];
						var dx = p1[0] - p2[0]; var dy = p1[1] - p2[1];
						var d = dx*dx+dy*dy; 
						metric += d;
					}
					// remember the best match
					if((best_score===undefined)||(best_score>metric)) {
						best_glyph=h; best_index=i; best_score=metric;
					}
				}
			}
		}
	}
	return { 
		glyph: best_glyph, index: best_index, metric: best_score 
	}
}
// break path into small regular segments to remove pen jitter
function path_regularize(path, dist) {
	var count = path.length; 
	if(count<=2) return path;
	// result points array
	var r = []; var error2 = 0; var dist2 = dist*dist;
	// add the start point
	var p = path[0].slice(0,2);  r.push(p);
	var d = 0;	var i = 0;
	while(i<count) {
		var np = path[i]; // next point along path
		var dp = [ np[0]-p[0], np[1]-p[1] ]; // delta vector
		var d2 = dp[0]*dp[0] + dp[1]*dp[1];  // distance squared
		// how far off? (matters for the last one).
		error2 = d2-dist2; 
		if(error2>0) {
			var nd = Math.sqrt(d2);
			// the next point is outside the distance. interpolate
			var inter = (dist - d) / (nd - d); // 0..1
			var ip = [ p[0]*(1-inter) + np[0]*inter, 
				       p[1]*(1-inter) + np[1]*inter ];
			r.push(ip); // this becomes the new start point
			p = ip; // loop and test again
		} else {
			i++; // the point is within the required distance.
		}
	}
	r.push(path[count-1].slice(0,2)); // add the final point
	return { 
		points: r, 
		error: dist - Math.sqrt(error2+dist2) 
	}
}

// interpolate a regular path into exact segment count
function path_interpolate(path, count) {
	if(path===undefined) return;
	var r = []; var pcount = path.length;
	for(var i=0; i<count; i++) {
		// next point index
		var npi = (pcount-1)*i/(count-1);
		// split it into integer and fraction
		var npi_int = Math.floor(npi);
		var npi_frac = npi - npi_int;
		// if the fraction is essentially zero...
		if(npi_frac<0.000000001) {
			// copy the indexed path point directly
			r.push(path[npi_int].slice(0));
		} else {
			// interpolate between path points
			var p1 = path[npi_int];	var p2 = path[npi_int+1];
			r.push([ p1[0]*(1-npi_frac)+p2[0]*npi_frac, 
				     p1[1]*(1-npi_frac)+p2[1]*npi_frac ]);
		}
	}
	return r;
}

// triangle orientation similarity between two glyphs
function tri_similarity(g,h,tri) {
	// get the edge vectors for both paths
	var ge = tri_edges(g, tri); var he = tri_edges(h, tri);
	// sum the cosine similarity
	var s = 0; for(var i=0; i<3; i++) s+=cos_angle(ge[i],he[i]);
	return s;
}

// extract the edge vectors for a path triangle
function tri_edges(path, tri) {
	// pull out the triangle vertexes
	var p0 = path[tri[0]]; 
	var p1 = path[tri[1]];
	var p2 = path[tri[2]];
	// create edge vectors
	var e0 = [ p0[0] - p1[0], p0[1] - p1[1] ];
	var e1 = [ p1[0] - p2[0], p1[1] - p2[1] ];
	var e2 = [ p2[0] - p0[0], p2[1] - p0[1] ];
	// return the set
	return [e0,e1,e2];		
}
// cosine angle between two vectors
function cos_angle(a,b) {
	return ( a[0]*b[0] + a[1]*b[1] )
	     / ( Math.sqrt(a[0]*a[0]+a[1]*a[1]) 
	         * Math.sqrt(b[0]*b[0]+b[1]*b[1]) );
}

// turn a path into a normalized triangle determinant matrix
function path_ntm(path, dist) {
	if(path===undefined) return { count:0 };
	var c = path.length;
	var c1 = c-1; var c2 = c-2; var c3 = c-3;
	if(dist===undefined) {
		// determine total path length by summing segment lengths
		dist = 0; var p1 = path[0];
		for(var i=1; i<c; i++) {
			var p2 = path[i]; 
			var xd = p2[0]-p1[0]; var yd = p2[1]-p1[1];
			var d = Math.sqrt(xd*xd+yd*yd);	dist += d;
		}
	}
	// all this to produce a single scaling number
	var scale = 4/(dist*dist);
	// produce the triangulation matrix
	var min; var max; var first = true; var tc = 0;
	// first axis is first point choice
	var a1 = [];
	for(var i=0; i<c2; i++) {
		var p1 = path[i]; var a2 = [];
		for(var j=i+1; j<c1; j++) {
			var p2 = path[j]; var a3 = [];
			// vector from first to second point
			var dx1 = p2[0] - p1[0]; var dy1 = p2[1] - p1[1];
			for(var k=j+1; k<c; k++) {
				var p3 = path[k];
				// vector from first to third point
				var dx2 = p3[0] - p1[0]; var dy2 = p3[1] - p1[1];
				// put them into a 2x2 matrix and take the determinant
				var md = dx1*dy2 - dx2*dy1;
				// normalize it, and put it in the final matrix
				var nd = md * scale;
				a3.push(nd); tc++;
				// include it in the metrics
				if(first) {
					min = nd; max = nd;	first = false;
				} else {
					min = Math.min(min,nd);	max = Math.max(max,nd);
				}
			}
			a2.push(a3);
		}
		a1.push(a2);
	}
	// do the metrics indicate the path is along a single line?
	var line = (Math.max(Math.abs(min),Math.abs(max))<0.004);
	// return result
	return {
		matrix: a1, path: path, line: line,
		count: tc, min: min, max: max
	}
}

// project path2 onto the co-ordinate system of path1
function path_project(path1, path2, tri) {
	// pull out the triangle vertexes from both paths
	var p10 = path1[tri[0]]; var p20 = path2[tri[0]]; 
	var p11 = path1[tri[1]]; var p21 = path2[tri[1]]; 
	var p12 = path1[tri[2]]; var p22 = path2[tri[2]];
	// construct an affine transformation matrix from each
	var t1 = [ p11[0]-p10[0], p11[1]-p10[1], p12[0]-p10[0], 
		         p12[1]-p10[1], p10[0], p10[1] ];
	var t2 = [ p21[0]-p20[0], p21[1]-p20[1], p22[0]-p20[0], 
		         p22[1]-p20[1], p20[0], p20[1] ];
	// affine mapping from t2 to t1 is t = t2^-1 * t1
	var t = svg_matrix_multiply( svg_matrix_inverse(t2), t1 );
	// run the points through the affine map
	var r = [];	
	for(var i in path2) r.push( svg_transform( t, path2[i] ) );
	return r;
}

// find a pivot point that selects the top [count] entries
function ntm_pivot_for(ntm, count, allow) {
	// start with full extents
	var ti = 1; var bi = 0;	var tc = ntm.count; bc = 0;
	// sanity check count
	if(tc<=count) return ti;
	// recursive subdivision
	while(true) {
		// compute a pivot in-between
		var mi = (ti + bi)/2;
		var mc = ntm_pivot_count(ntm,mi);
		// which way do we recurse?
		if(mc>=(count-allow) && mc<=(count+allow)) return mi;
		if(mc>count) { bi = mi; bc = mc; } 
			    else { ti = mi; tc = mc; }
		// stop if the difference gets too small.
		if((ti-bi)<0.00000001) return ti; // exclude ties
	}
}

// count all the triangles larger than [value]
function ntm_pivot_count(ntm, value) {
	var count = 0;
	ntm_each(ntm, function (i,j,k,v) {
		if(Math.abs(v)>=value) count++;
	});
	return count;
}

// get the set of all the triangles larger than [value]
function ntm_pivot_set(ntm, value) {
	var set = [];
	ntm_each(ntm, function (i,j,k,v) {
		if(Math.abs(v)>=value) set.push([i,j,k,v]);
	});
	return set;
}

// retrieve a triangle entry from the matrix
function ntm_entry(ntm, i, j, k) {
	try { return ntm.matrix[i][j-i-1][k-j-1]; } 
	catch(e) { return 0; }
}

// call a function for each triangle in the matrix
function ntm_each(ntm, fn) {
	var a1 = ntm.matrix; var c1 = a1.length;
	for(var i=0; i<c1; i++) {
		var a2 = a1[i]; var c2 = a2.length;
		for(var j=0; j<c2; j++) {
			var a3 = a2[j]; var c3 = a3.length;
			for(var k=0; k<c3; k++) fn( 0+i, 1+i+j, 2+i+j+k, a3[k] );
		}
	}
}

// multiply two svg transformation matrices
function svg_matrix_multiply(a, b) {
	return [
	    a[0]*b[0] + a[1]*b[2], a[0]*b[1] + a[1]*b[3],		
	    a[2]*b[0] + a[3]*b[2], a[2]*b[1] + a[3]*b[3],		
	    a[4]*b[0] + a[5]*b[2] + b[4],		
	    a[4]*b[1] + a[5]*b[3] + b[5]	
	];
}

// inverse of svg transformation matrix 
function svg_matrix_inverse(m) {
	var a = m[0]; var c = m[2]; var e = m[4]; 
	var b = m[1]; var d = m[3]; var f = m[5]; 
	var det = a*d-c*b;
	return [ d/det, -b/det, -c/det, a/det, 
		     (c*f-d*e)/det, -(a*f-b*e)/det ];
}

// multiply vector v by matrix a. 
function svg_transform(a, v) {
	return [ v[0]*a[0] + v[1]*a[2] + a[4],
		      v[0]*a[1] + v[1]*a[3] + a[5] ];
}

\end{lstlisting}
\end{multicols}

\section{Further Improvements}
As well as the obvious code optimizations that would improve speed, several complications can help the algorithm in practice. When recognizing glyphs as they are being input in real-time, we can exploit work done previously:
\begin{itemize}
\item Append to the regularized path as new input is received, rather than rebuilding it from scratch. Just remember that the regularized path segments are exactly the same distance apart, which means they are \emph{not} pixel-aligned.
\item If an established path is only extended by a few pixels, it's NTM (and therefore it's largest triangle ordering) will be largely unchanged. The previous sort order/pivot point is likely to be close to correct, and is a good starting point for an incremental algorithm.
\item The pivot search can be improved in dozens of ways: such as reducing the list during some iterations, (by sorting the list into `above' and `below' lists, counting them, and sequestering one side) taking averages or histograms to use as pivots instead of pure recursive approximation, or even random sampling. 
\item If the triangle matrices become infeasible, then we can get very good approximations by dropping to a lower resolution. We might even use the low-res information to sparsely render a high-resolution matrix.
\end{itemize}

\section{Observations}
Why does it work so well? For that reason, why does it work at all? We seem to spend a lot of computation time looking at the input glyph for triangles, and very little effort comparing glyphs geometrically, which would seem to be the whole point.

There is a statistical method called ``Principal Component Analysis'' (PCA) which takes a cloud of point samples and assumes it is a (usually Gaussian) distribution around a mean in the form of a squashed ovoid. (In contrast, The usual statistical mean and standard deviation tests assume a simple `spherical' distribution around the average.) A 2-D PCA analyzes the cloud of samples and returns two vectors which represent the cloud's \emph{long axis} and \emph{short axis}, both in direction and size. Once these principal components are known, the sample cloud can be mapped onto other set for comparison.

One method of calculating the PCA is performing \emph{singular value decomposition} to deconstruct the full covariance matrix as a series of simpler transformations. The algorithm essentially tries to force the squashed ovoid back into a standard distribution through a series of affine transforms which are generated from internal features of the sample points. The transform which does this best naturally contains the principal axes.

This sounds very similar in concept to Squiggle, and for good reason. The largest-area triangle search is looking for the biggest pair of mostly-orthogonal axes which can be made from the input points; it's essentially looking for principal components, as important and essential a numeric property as the average or standard deviation. The results are therefore useful for comparing against \emph{all} templates; we don't even have to know what they are yet!

\section{Conclusion}
Squiggle fulfills all the requirements to be a practical glyph recognizer that can handle large dynamic libraries of arbitrary user-programmed template glyphs. The major computational step of the algorithm is fixed time and is it's own natural approximation at lower resolutions, so performance tradeoffs can make the algorithm suitable for a wide range of hardware. The algorithm is fast, robust, numerically stable, and handles it's own worst cases elegantly. There are no bad glyphs that give the recognizer trouble, and it can gracefully tune it's approximations to cope with changes in available resources. It's characteristics make it almost ideal as a part of a real-time computer input system using pen-like pointer devices.

\end{document}